\documentclass[a4paper]{article}
\usepackage{amssymb}
\usepackage{fontenc}
\usepackage{times}
\usepackage{mathptmx}
\usepackage{graphicx}
\usepackage[]{txfonts}
\usepackage{enumerate}

\newcommand{\dsfrac}[2]{\displaystyle{\frac{#1}{#2}}}

\def\gph{\Gamma_{\rm ph}}

\title{Numerical simulations of the internal shock model in magnetized relativistic jets of blazars}

\author{Jesus Rueda-Becerril
  and Petar Mimica and Miguel A. Aloy \\
  {\footnotesize Departamento de Astronom\'ia y Astrof\'isica, Universidad de Valencia, E-46100 Burjassot (Valencia), Spain}\\
  {\footnotesize E-mail: \texttt{jesus.rueda@uv.es}, \texttt{petar.mimica@uv.es}, \texttt{miguel.a.aloy@uv.es}}}

\date{}

\begin{document}

\maketitle

\begin{abstract}
  The internal shocks scenario in relativistic jets is used to explain 
  the variability of the blazar emission. Recent studies have shown 
  that the magnetic field significantly alters the shell collision 
  dynamics, producing a variety of spectral energy distributions and 
  light-curves patterns. However, the role played by magnetization in 
  such emission processes is still not entirely understood. In this 
  work we numerically solve the magnetohydodynamic evolution of the 
  magnetized shells collision, and determine the influence of the 
  magnetization on the observed radiation. Our procedure consists in 
  systematically varying the shell Lorentz factor, relative velocity,
  and viewing angle. The calculations needed to produce the whole 
  broadband spectral energy distributions and light-curves are 
  computationally expensive, and are achieved using a high-performance 
  parallel code.
\end{abstract}

\section{Introduction}
Blazars are a class of radio-loud active galactic nuclei (AGNs) whose
jets are pointing near the line of sight to the observer
\cite{urr1995}, and are known for showing the most rapid variability
of all AGNs. Extensive observations indicate that the jets of blazars
are relativistic. Their most remarkable characteristic is the presence
of flares in the X- and $\gamma$-ray bands, usually wih a duration of
the order of a few hours. The internal shock (IS) model
\cite{ree1994}, invoked to explain this variability
\cite{mim2004,mim2005, spa2001}, is an idealized model where an
intermittently working central engine ejects shells of magnetized
plasma which collide due to their velocity differences. As a
consequence internal shocks are formed, accelerating particles at the
shock fronts and producing the non-thermal, highly variable radiation.

Our long-term objective for is the study of the influence of the
magnetic field on the observed emission using numerical simulations.
In \cite{mim2012} we studied a large number of shell collisions with
different magnetization levels. In \cite{rue2014} we focused on a
limited number of parameters such as the observer viewing angle,
$\theta$, the bulk Lorentz factor of the slower shell, $\Gamma_R$, and
the relative Lorentz factor, $\Delta g := (\Gamma_L / \Gamma_R) - 1$,
where $\Gamma_L$ is the bulk Lorentz factor of the faster shell. In
the present work we summarize some of the results studied therein.

We describe the common numerical setup in Sec.~\ref{num-setup}. The results
are shown in Sec.~\ref{results} and discussed in
Sec.~\ref{conclusions}.

\section{Numerical Setup}
\label{num-setup}
A modified version of the code SPEV \cite{mim2009} was employed to
calculate the non-thermal emission from ISs. We assume a cylindrical
shell geometry and perform all the calculations in the rest frame of
the shocked fluid (RFSF). The shell interaction was simplified as a
one-dimensional Riemann problem and focus our resources on a more
detailed treatment of the non-thermal radiation.  Our method consists
of three steps that we sum up in the following paragraphs.

\emph{Solution of the Riemann problem.} Employing an exact
relativistic magnetohydrodynamics (RMHD) Riemann solver we determine
the properties of the internal shock waves. We follow the procedure
described in \cite{mim2012} to set-up the shells and to extract the
information needed for the steps 2 and 3.

\emph{Non-thermal particles transport and evolution.} The particles
are injected behind the shock fronts following the prescription of
\cite{bot2010,jos2011,mim2012}. We assume that a fraction of the
thermal electrons are accelerated into a power-law distribution at
high energies, and that their energy density is a fraction of the
internal energy density of the shocked fluid. In the RFSF the shocks
are propagating away from the initial discontinuity, injecting and
leaving non-thermal particles behind. We evolve the energy
distribution of non-thermal electrons taking into account synchrotron
and inverse-Compton (IC) losses. See \cite{mim2012} for more details.

\emph{Radiative transfer.} The total emissivity at each point is
assumed to be a combination of the following emission processes: (1)
synchrotron radiation, (2) IC upscattering of an internal (SSC) and an
external radiation field (EIC).

\section{Results}
\label{results}
We group our models according to the initial shell magnetization,
$\sigma := B^{2} / 4\pi \rho \Gamma^2 c^2$. Hereafter the subscripts
$L$ and $R$ will denote left (faster) and right (slower) shells,
respectively. We compute the spectra for a typical source located at
$z = 0.5$. In the rest of this section we will present some of the
final SEDs resulting from our simulations. A larger collection is
shown in \cite{rue2014}.  The SED of each model has been averaged over
the time interval $0-10^{6}\,$s.

\subsection{Weakly-magnetized models}
\label{sec:weak}
The SEDs computed for the models with $\sigma_L = \sigma_R = 10^{-6},
\Gamma_R = 10$, $\theta = 5^\circ$ and varying $\Delta g$, are shown
in the upper left panel of Fig.~\ref{fig:W-G10-T5}. The spectra show
that with increasing $\Delta g$ the IC component also increases, up to
three orders of magnitude. In order to see the effects on each
emission process, the synchrotron, SSC and EC components for $\Delta g
= 0.5, 2.0$ are shown as dashed, dot-dashed and dot-dot-dashed lines,
respectively. While the three spectral components (synchrotron, SSC
and EC) have approximately the same order of magnitude for $\Delta g =
0.5$, for $\Delta g = 2.0$ the SSC is almost two orders of magnitude
more luminous than the other two.

\subsection{Moderately-magnetized models}
\label{sec:moderate}
The SEDs of the family of models $\sigma_L = \sigma_R = 10^{-2},
\Delta g = 1.0$ and $\theta = 5^\circ$ are presented in the upper
right panel of Fig.~\ref{fig:M-D1p0-T5}. As in the previous section,
SSC and EC components are shown as dashed, dot-dashed and
dot-dot-dashed lines, respectively, but now for the models with
$\Gamma_R = 10, 17, 25$. The synchrotron component for $\Gamma_R = 10$
is $\simeq 20$ times brighter than the SSC one, in contrast to the EC
which is $100$ times dimmer. For $\Gamma_R = 25$ the EC is of the same
order of magnitude of SSC and synchrotron. The latter two decrease one
order of magnitude between the models with $\Gamma_R = 10$ and
$\Gamma_L$, while the EC grows by almost one order of magnitude. This
follows from the fact that the number of electrons and the
comoving magnetic field strength decrease with the increasing
$\Gamma_R$ \cite{mim2012}, which means that there are less synchrotron
photons and less electrons which can scatter them in the SSC process.
We model the Klein-Nishina cutoff as a sharp cutoff, and its effect
can be seen at $\simeq 10^{23}\,$Hz, where all the EC spectra
coincide. From the inset we can see that there is no significant
change in the flux of $\gamma$-photons, although the spectral
index softens for increasing $\Gamma_R$.

\subsection{Strongly-magnetized models}
\label{sec:strong}

\begin{figure}
  \centering
  \includegraphics[width=6.8cm,clip]{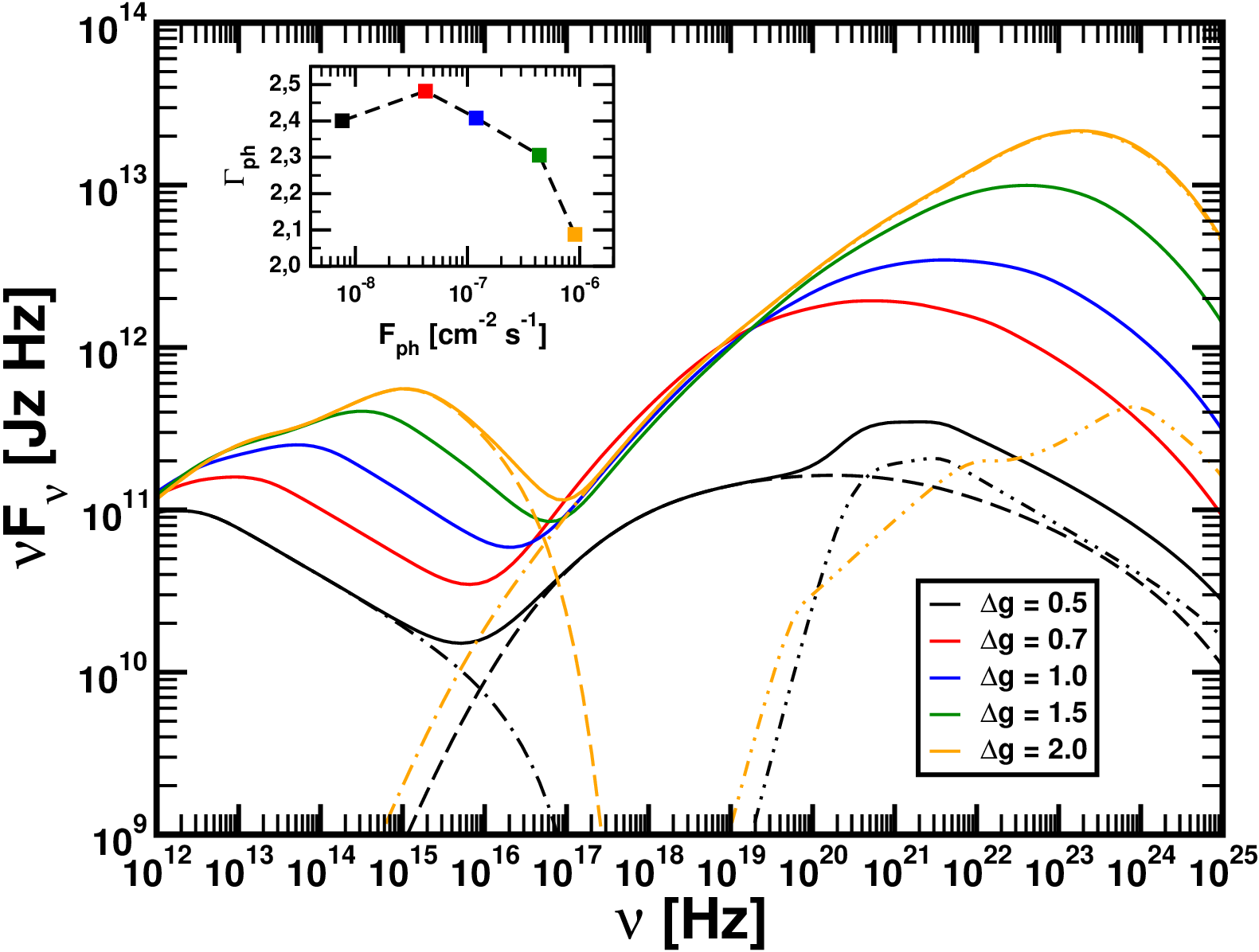} 
  \ \hspace{0.16cm}
  \includegraphics[width=6.8cm,clip]{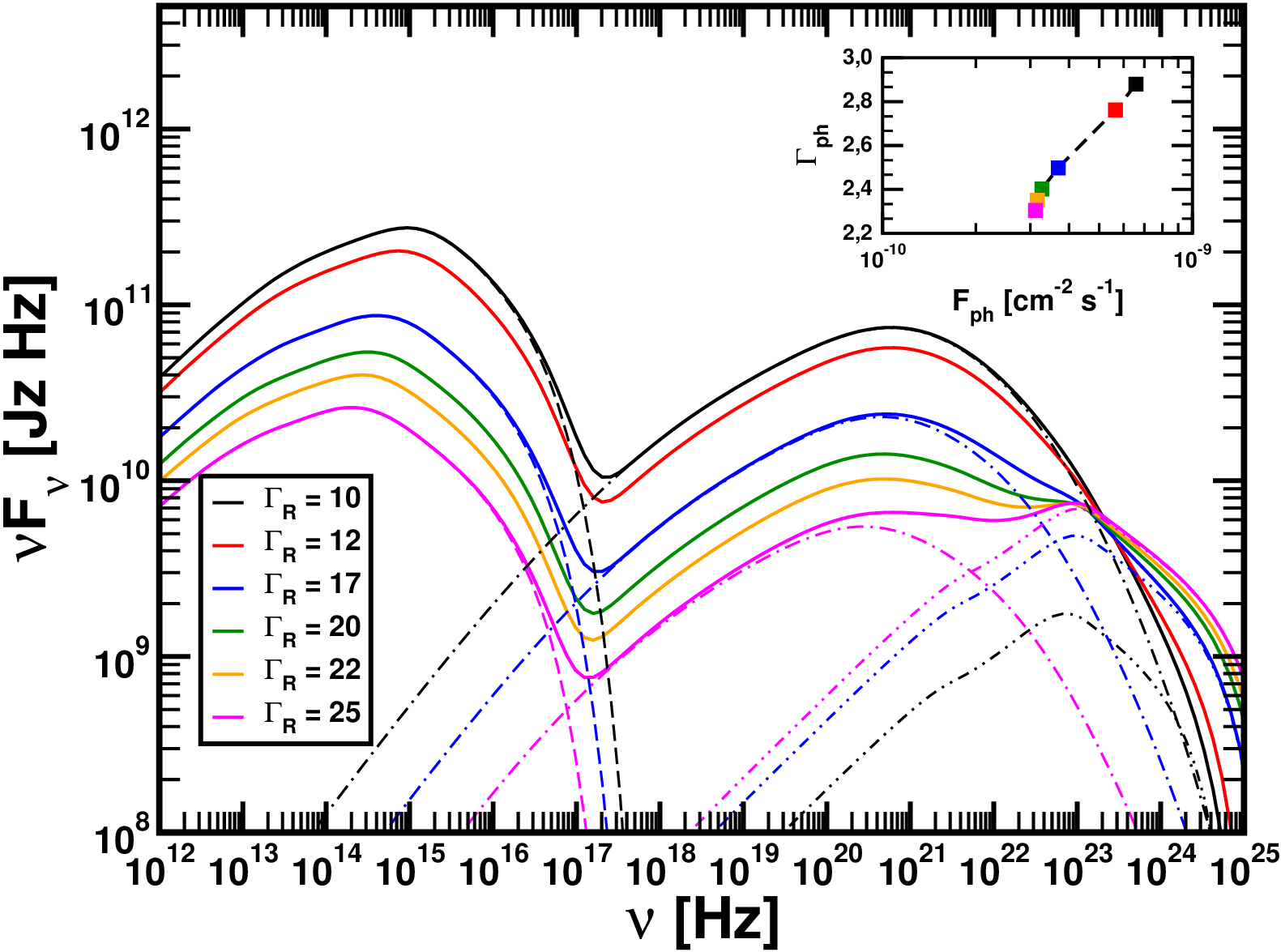}
   \ \hspace{0.16cm}
  \includegraphics[width=6.8cm,clip]{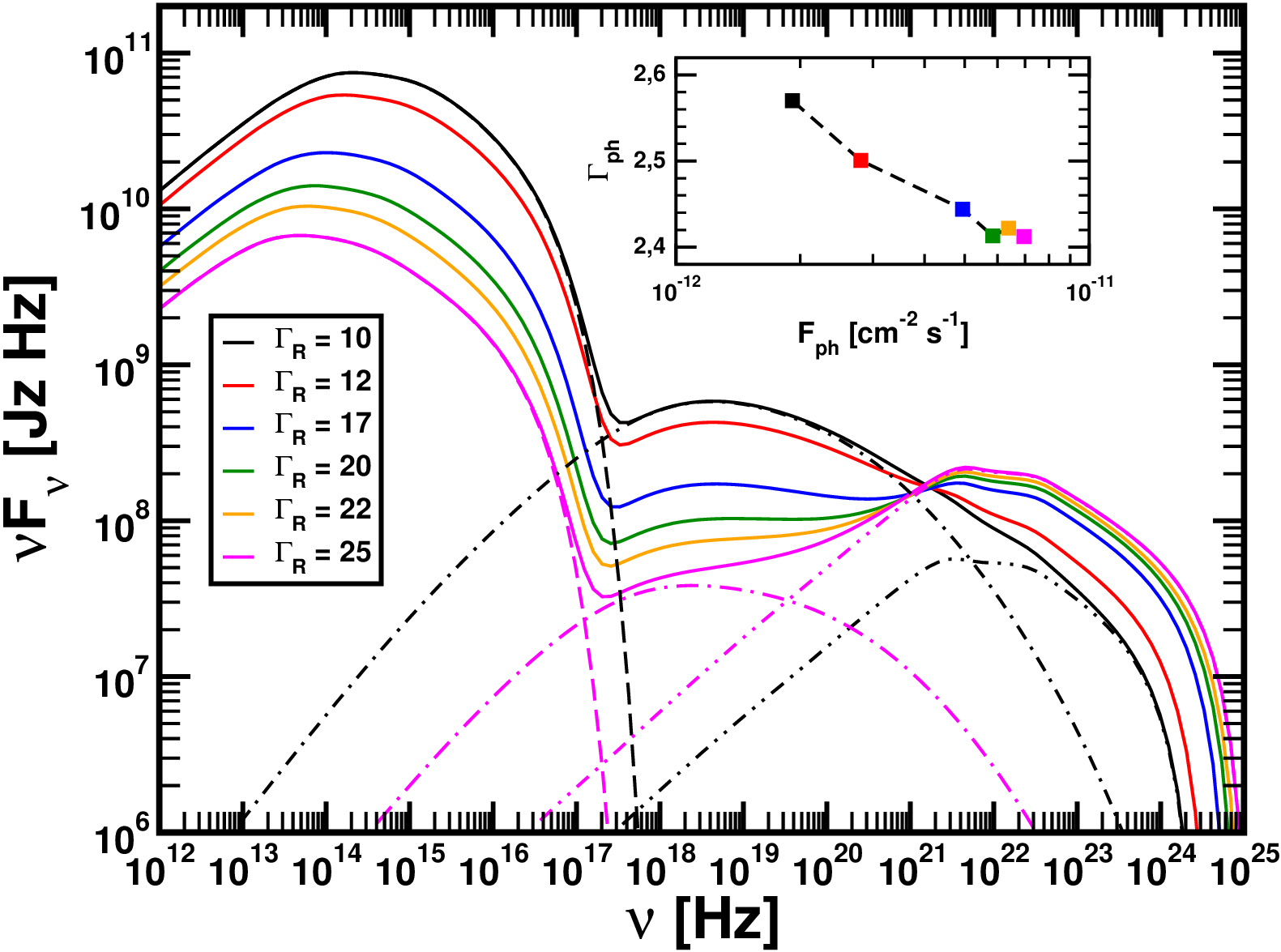}
  \caption{Upper left panel: Averaged spectra resulting from collisions of weakly magnetized shells ($\sigma_L =
    \sigma_R = 10^{-6}$). Upper right panel: Same as the upper left panel, but for moderately magnetized shells
    ($\sigma_L = \sigma_R = 10^{-2}$) and varying $\Gamma_{R}$. Lower panel: Same as upper right panel, but for strongly
    magnetized shells ($\sigma_L = 1, \sigma_R = 0.1$).  The inset in each panel shows the photon spectral slope
    $\Gamma_{ph}$ as a function of the photon flux $F_{ph}$ in the $\gamma$-ray band (see Sec.\,3.2).  Colors of the
    points correspond to the line colors in the main plot. For selected models, the synchrotron, SSC and EIC
    contributions (dashed, dot-dashed and dot-dot-dashed lines, respectively) are shown.}
  \label{fig:W-G10-T5}
   \label{fig:M-D1p0-T5}
   \label{fig:S-D1p0-T5}
 \end{figure}

The third family consists of the strongly magnetized models
where $\sigma_L = 1$ and $\sigma_R = 0.1$. The corresponding SEDs of
this series of models (lower panel of Fig.~\ref{fig:S-D1p0-T5} are
analogous to the models of Sec.~\ref{sec:moderate}. As we can see, for
$\Gamma_R = 10$ the synchrotron component is $\simeq 100$ times
brighter than the IC. For $\Gamma_R = 25$ this difference decreases to
one order of magnitude. Once again, EIC component rises with rising
$\Gamma_R$, to the point in which it begins to be comparable to the
synchrotron component, developing an IC hump. The EIC spectra
interestect in a single point due to our treatment of the
Klein-Nishina cutoff. In the inset we can see that the flux of
$\gamma$-ray photons does not change appreciably in this family of
models. Table~\ref{tab:Smicro1} lists a number of physical parameters
in the reverse shock (RS) of the present models. We see that the
electrons in the RS of the strongly magnetized models are
fast-cooling. In fact, for $\Delta g = $1.5 the injected electron
spectrum is almost mono-energetic.

\begin{table*}
  \centering
  \begin{tabular}{rrrrcrrrr}
    \hline \hline 
    $\Delta g$ & $\Gamma$ & $r_r$ & $\dsfrac{B_{r}}{1{\rm G}}$ &
    $\dsfrac{Q_{r,11}}{{\rm cm}^{-3} {\rm s}^{-1}}$ &
    $\dsfrac{\gamma_{1r}}{10^2}$ & $\dsfrac{\gamma_{2r}}{10^4}$ &
    $\dsfrac{t'_{crr}}{10^3 {\rm s}}$ &
    $\dsfrac{\gamma_{cr}}{\gamma_{1r}}$ \\
    \hline
    $ 0.5$ & $   12.7$ & $   1.26$ & $  53.51$ & $      0.11$ & $
    0.66$ & $   0.64$ & $   34.6$ & $   0.12$ \\
    $ 0.7$ & $   12.8$ & $   1.46$ & $  54.72$ & $      1.03$ & $
    2.29$ & $   0.63$ & $   34.1$ & $   0.03$ \\
    $ 1.0$ & $   13.0$ & $   1.75$ & $  55.84$ & $      7.33$ & $
    7.25$ & $   0.62$ & $   33.6$ & $   0.01$ \\
    $ 1.5$ & $   13.2$ & $   2.22$ & $  56.63$ & $     68.00$ & $
    26.38$ &$   0.62$ & $   32.9$ & $  0.003$ \\
    $ 2.0$ & $   13.3$ & $   2.67$ & $  56.82$ & $ 112900.75$ & $
    61.68$ & $   0.62$ & $   32.5$ & $ 0.001$ \\
    \hline
  \end{tabular}
  \caption{Physical parameters in the RS shocked region for the
    family of strongly magnetized shells with different $\Delta
    g$. The bulk Lorentz factor of both shocked regions is denoted by
    $\Gamma$, while $r$, $B$, $Q$, $\gamma_1$ and $\gamma_2$ denote
    its compression ratio, comoving magnetic field, comoving number of
    electrons injected per unit volume and unit time, and lower and
    upper cutoffs of the injected electrons in the RS (see Eq.~11 of
    MA12). $Q_{11} = Q\times 10^{-11}$. $t'_{cr} := \Delta r' / (c
    |\beta'|)$ is the shock crossing time, where $\Delta r'$ and
    $\beta'$ are the shell width and the shock velocity in the frame
    moving with the  contact discontinuity separating both shocks
    (section 2 of MA12). $\gamma_c := \gamma_2 / (1 + \nu_0 \gamma_2
    t'_{cr})$ is the cooling Lorentz factor of an electron after a
    dynamical time scale (shock crossing time). $\nu_0:=(4/3)c\sigma_T
    (u'_B + u'_{\rm ext}) / (m_e c^2)$ is the cooling term, where
    $\sigma_T$ is the Thomson cross section and the primed quantities
    are measured in the comoving frame. When $\gamma_c / \gamma_1 \gg
    (\ll) 1$ the electrons are slow (fast) cooling. Note that the
    $Q_{11}$ for $\Delta g = 2.0$ is much larger than $Q_{11}$ of the
    other models because $\gamma_1\simeq\gamma_2$.}
  \label{tab:Smicro1}
\end{table*}

\subsection{$\gamma$-rays spectral slope}
\label{sec:spectralslope}
\begin{figure}
  \centering 
  \includegraphics[height=5cm,clip]{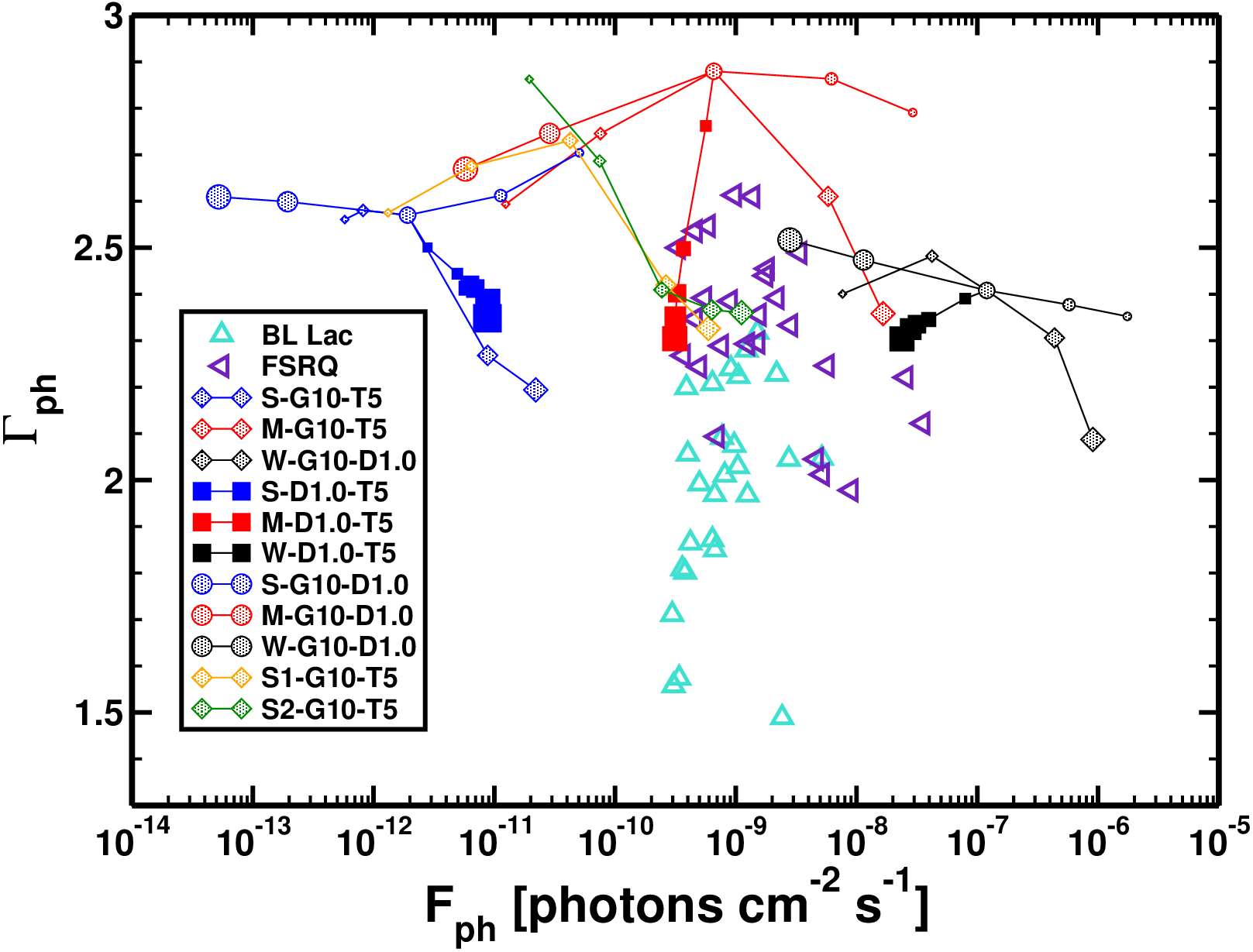}
  \ \hspace{0.16cm}
  \includegraphics[height=5cm,clip]{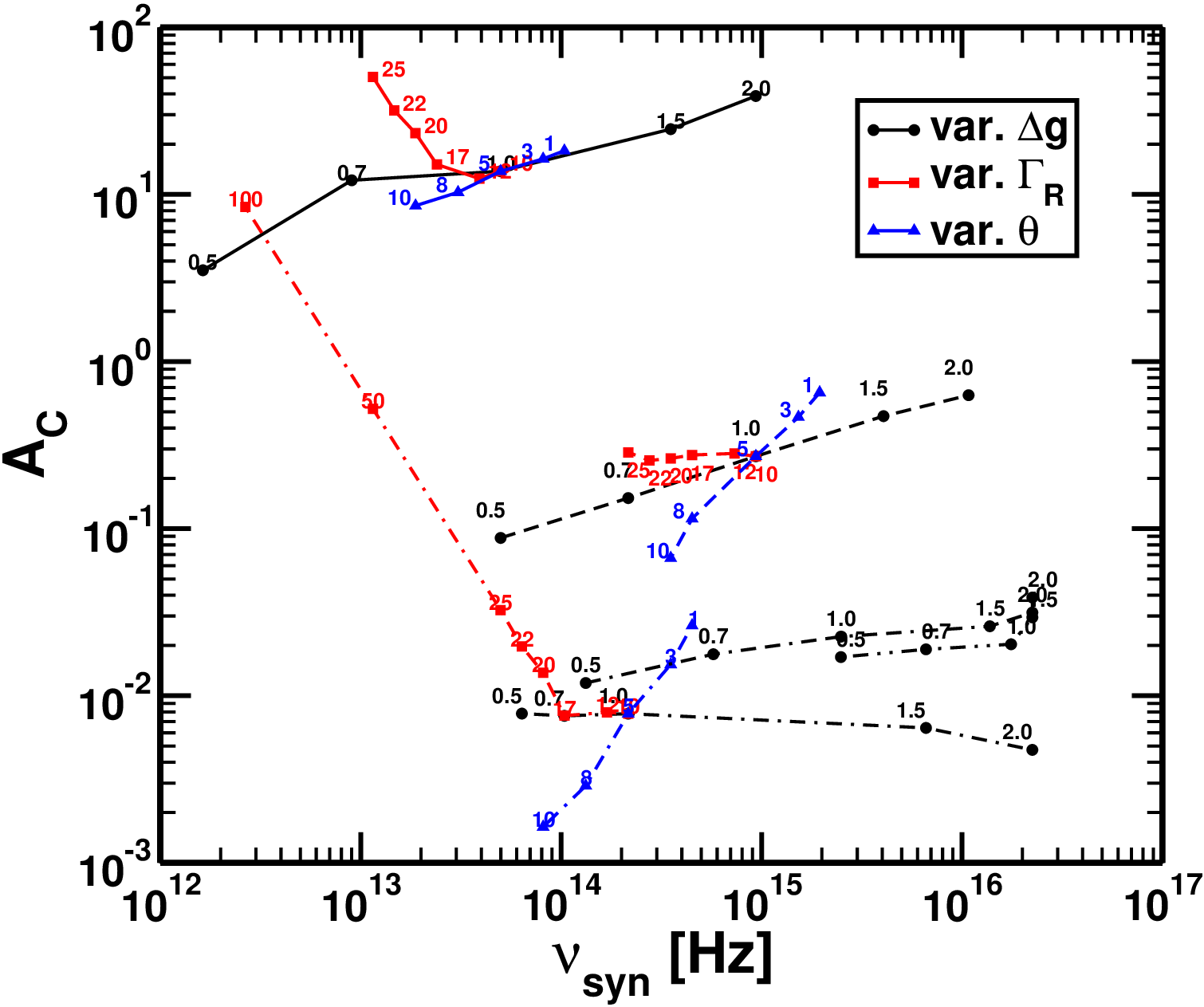}
  \caption{Left panel: Spectral slope $\Gamma_{ph}$ for the photon
    energies $>200\,$MeV as a function of the photon flux for energies
    $>100$ MeV \cite{abd2009}. The symbols joined by lines represent
    our numerical models, while cyan and magenta triangles represent
    BL Lacs and FSRQs at redshift $z\simeq 0.5$ from 2LAC
    \cite{ack2011}. Right panel: Compton dominance $A_C$ as a function
    of the synchrotron peak frequency $\nu_{\rm syn}$ for all the
    models studied in \cite{rue2014}.}
  \label{fig:spectral-slope}
  \label{fig:compton-dominance}
\end{figure}

A linear least-squares algorithm is used to deduce the $\gamma$-ray
spectral slope $\Gamma_{ph}$. Due to the fact that we are not modeling
the Klein-Nishina part of the spectrum, we only performed the
calculations of $\Gamma_{ph}$ for those models that do not show a
large drop-off in the photon flux. In Fig.~\ref{fig:spectral-slope} we
show $\Gamma_{ph}$ as a function of the photon flux for energies $>
0.2\,$GeV, where $F_{ph}$ is the photon flux for photon energies $>
0.1$ GeV \cite{abd2009}. We compare our results with sources found in
2LAC \cite{ack2011}, restricting the comparsion to sources with $0.4
\leq z \leq 0.6$. In Fig.~\ref{fig:spectral-slope} we see that weakly
and moderately magnetized models overlap with the observations, with
more weakly than moderately magnetized models falling within the
observed part of the parameter space.

\subsection{Compton dominance $A_C$}
Another way to correlate magnetization with observed properties can be
found by representing the Compton dominance $A_C$ as a function of the
ratio of IC-to-synchrotron peak frequencies $\nu_{IC} /
\nu_{syn}$. Models with intermediate or low magnetization occupy a
range of $A_C$ roughly compatible with observations \cite{fin2013},
while the strongly magnetized models tend to have values of $A_{C}$
that are probably uncompatible with those observed in actual sources,
unless collisions in blazars happen at much larger Lorentz factors
than currently thought.

\section{Conclusions}
\label{conclusions}
We have shown that the SEDs of FSRQs and BL Lacs strongly depend on
the magnetization of the emitting plasma. Our models predict a more
complex phenomenology than what is currently supported by the
observational data. A conservative conclusion is that the observations
restrict the probable magnetization of the colliding shells that take
place in actual sources to, at most, moderate values (i.e., $\sigma
\lesssim 10^{-1}$), and if the magnetization is large, with variations
in magnetization between colliding shells which are smaller than a
factor $\sim 10$. We find that FSRQs have observational properties
corresponding to models with negligible or moderate magnetic fields.
BL Lacs with moderate peak synchrotron frequencies $\nu_{syn}\lesssim
10^{16}\,$Hz and Compton dominance parameter $0.1\gtrsim A_C \gtrsim
1$ display properties that can be reproduced with models with moderate
and uniform magnetization ($\sigma_L=\sigma_R=10^{-2}$). We 
find that a fair fraction of the {\em blazar sequence} can be
explained in terms of the intrinsically different magnetization of the
colliding shells. We observe that the change in the photon spectral
index ($\gph$) in the $\gamma-$ray band can be a powerful
observational proxy for the actual values of the magnetization and of
the relative Lorentz factor of the colliding shells. Values $\gph
\gtrsim 2.6$ result in models where the flow magnetization is $\sigma
\sim 10^{-2}$, whereas strongly magnetized shell collisions
($\sigma>0.1$) as well as weakly magnetized models may yield $\gph
\lesssim 2.6$.

\section*{Acknowledgements}
We acknowledge the support from the European Research Council (grant
CAMAP-259276), and the partial support of grants AYA2013-40979-P,
CSD2007-00050 and PROMETEO-2009-103, and to the Consejo Nacional de
Ciencia y Tecnolog\'{\i}a (MEXICO) for a doctoral fellowship to
J.M.R.B.

\end{document}